\documentstyle[epsfig]{aipproc}
\newcommand{\bm}{\bibitem}

\def\be {\begin{equation}}
\def\ee {\end{equation}}
\def\bea {\begin{eqnarray}}
\def\eea {\end{eqnarray}}

\def\2l {\frac{{f_i}}{(2\lambda + 1)}}

\def\var {\bf}

\begin{document}
\title{Meson Mixing and Dilepton Production in Heavy Ion Collisions}
\author{A.K. Dutt-Mazumder, C. Gale\thanks{Speaker.}, and O. Teodorescu}
\address{ Physics Department, McGill University\\ 3600 University St., 
Montreal, Quebec H3A 2T8, Canada\\}
\maketitle

\begin{abstract}
We study the possibility of $\rho-a_0$ mixing via N-N excitations in 
dense nuclear matter. This mixing is found to induce a peak in the
dilepton spectra at an invariant mass equal to that of the $a_0$.
We calculate the cross section for dilepton production through 
mixing and we compare its size with that of
$\pi-\pi$ annihilation. In-medium 
masses and mixing angles are also calculated. Some preliminary 
results of the mixing effect on the dilepton production rates at finite temperature are also presented.
\end{abstract}
\vspace{0.3 cm}

%
\vspace{.2 cm}

Electromagnetic radiation, especially lepton pairs,  constitutes a
class of  valuable 
probes in the context of heavy ion collisions.
This owes to the fact that the leptons couple 
to hadrons via vector mesons and therefore hadronic processes involving 
$\ell^+ \ell^-$ in the final channel are expected to reveal 
their properties in
the dilepton spectra. 
Furthermore, the $\ell^+ \ell^-$ pairs suffer minimum final state
interactions and are thus likely to bring information to the detectors 
essentially unscathed. We will only consider dielectrons in this work.

Several experiments have measured, or are planning to measure,  the lepton 
pairs produced in
nucleus-nucleus collisions. They have been carried out by  
the DLS at LBL \cite{dls}, and by 
HELIOS \cite{hel} and 
CERES \cite{ceres} at CERN. Two new initiatives that will focus on
electromagnetic probes will be PHENIX at RHIC \cite{phenix} and HADES at 
GSI \cite{hades}.  
The density-dependent characteristics of vector mesons can also be
highlighted through experiments performed at TJNAF \cite{tjnaf}.


We explore here the possibility of $\rho$-$a_0$ mixing via n-n 
excitations in nuclear matter.
This is a pure density-dependent effect 
and is forbidden in free space on account of Lorentz symmetry. We show 
that such a mixing opens up a new channel in dilepton production 
and will modify the spectrum in the $\phi$ mass region. The details of
the calculation will only be sketched here. The interested reader is invited
to consult \cite{tdmg}.

The interaction Lagrangian we will use can written as 
\bea
{\cal L}_{int}& = & g_\sigma {\bar \psi}\phi_\sigma \psi + 
          g_{a_0} {\bar \psi}\phi_{a_0,a}\tau^a \psi
          + g_{\omega NN}{\bar{\psi}} 
\gamma _\mu\psi\omega^\mu \nonumber \\
          & +&  
 g_{\rho} [{\bar{\psi}} \gamma _\mu \tau^\alpha
 \psi + \frac{\kappa _\rho}{2m_n}{\bar{\psi}}
    \sigma_{\mu\nu}\tau^\alpha \partial ^\nu] \rho^\mu_\alpha\ ,
\eea                                                 
where $\psi$, $\phi_\sigma$, $\phi_{a_0}$, $\rho$ and $\omega$ correspond
to nucleon, $\sigma$, $a_0$ , $\rho$ and $\omega$ fields, and $\tau_a$ 
is a  Pauli matrix. The values used for the coupling parameters are obtained 
from Ref. \cite{mach89}.

The polarization vector through which the $a_0$ couples to $\rho$
via the n-n loop is given by 
\bea
\Pi_ \mu (q_0,|{\vec q}|) &=& 2i g_{a_0} g_\rho \int \frac{d^4k}{(2\pi)^4}
    \mbox{Tr}[G(k) \Gamma_\mu G(k+q)]. \label{pim}\ ,
\eea
where 2 is an isospin factor and the vertex for $\rho$-nn
coupling is
\bea
\Gamma_\mu=\gamma_\mu - \frac{\kappa_\rho}{2m_n}\sigma_{\mu\nu}q^\nu\ .
\label{vertex}
\eea
$G(k)$ is the in-medium nucleon propagator \cite{sewal}. 

With the evaluation of the trace and after a little algebra Eq.~(\ref{pim}) could
be cast into a suggestive form: 
\bea
\Pi_\mu(q_0,|q|)=\frac{g_\rho g_{a_0}}{\pi^3} 2q^2 (2m_n^*-
\frac{\kappa q^2}{2 m_n})
\int_0^{k_F}\frac{d^3k}{E^*(k)}
\frac{k_\mu - \frac{q_\mu}{q^2}(k\cdot q)}{q^4 - 4 (k\cdot q)^2}\ .
\label{mixamp}
\eea
It easily can be seen that it obeys current conservation: 
$q^\mu\Pi_\mu=0=\Pi_\nu q^\nu$. This implies that only one component 
of $\Pi_\mu$ is independent. In fact, only the longitudinal component of the 
$\rho$ meson couples to the scalar meson while the transverse
mode remains unaltered. 

In presence of mixing the combined meson propagator might be written
in a matrix form where the dressed propagator would no longer be diagonal: 
\bea
{\cal D} = {\cal D}^0 + {\cal D}^0\Pi{\cal D} \hspace*{15mm} 
{\cal D}^0 = \pmatrix{
    D^0_{\mu \nu} & 0\cr
    0 & \Delta_0
}\ . 
\label{dressprop}
\eea
The noninteracting propagator for the $a_0$
and $\rho$ are given respectively by
\bea
{\Delta_0(q) = \frac{1}{q^2 - m_{a_0}^2 + i\epsilon}}\ \ , \hspace*{15mm}
{D^0_{\mu \nu}(q) = \frac{-g_{\mu \nu}+ {\frac{q_\mu q_\nu}{q^2}}}{
  q^2 - m_\rho^2 + i\epsilon}} \ .
\label{freerho}
\eea

The mixing is characterised by the polarization matrix which contains
non-diagonal elements 
\be
\Pi = \pmatrix{
    \Pi^\rho_{\mu \nu}(q) & \Pi_\nu(q)\cr
    \Pi_\mu(q) & \Pi^{a_0}(q)}\ .
\label{pol}
\ee
$\Pi^{a_0}$ and $\Pi^\rho_{\mu\nu}$ refer to
the diagonal self-energies of the $a_0$ and $\rho$ meson induced by 
the n-n polarization. 

The $\rho$ meson being a vector, one can write the longitudinal and 
transverse polarization as $\Pi_L= - \Pi_{00} + \Pi_{33} $ and 
$\Pi_T=\Pi_{11}=\Pi_{22}$. 
To determine the collective modes and their dispersion relation, one 
can define the
dielectric function and look for its zeroes \cite{chin77}.
%
%
\begin{figure} [htb!]
\begin{center}
\epsfig{file=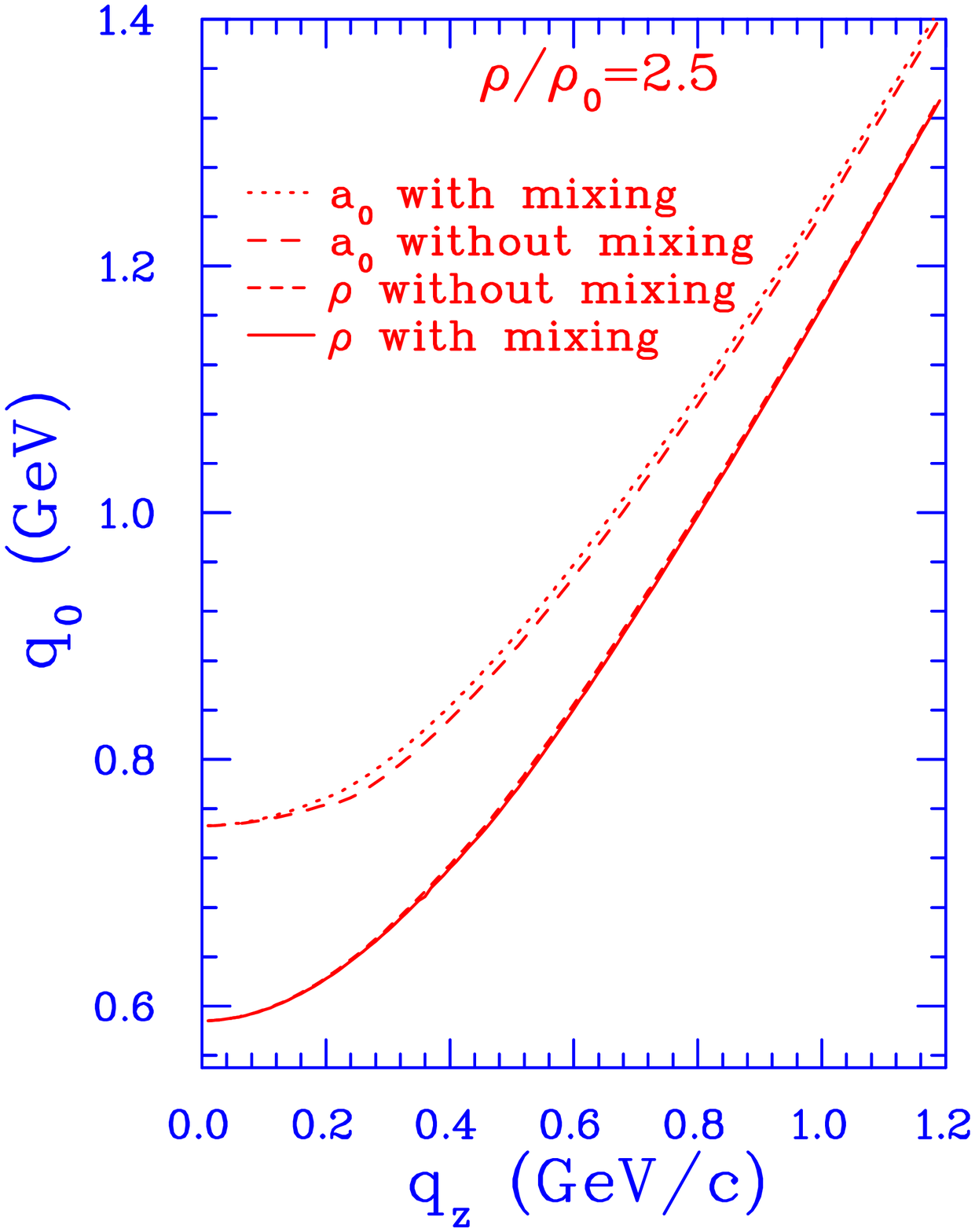,height=6.0cm,angle=0}
\epsfig{file=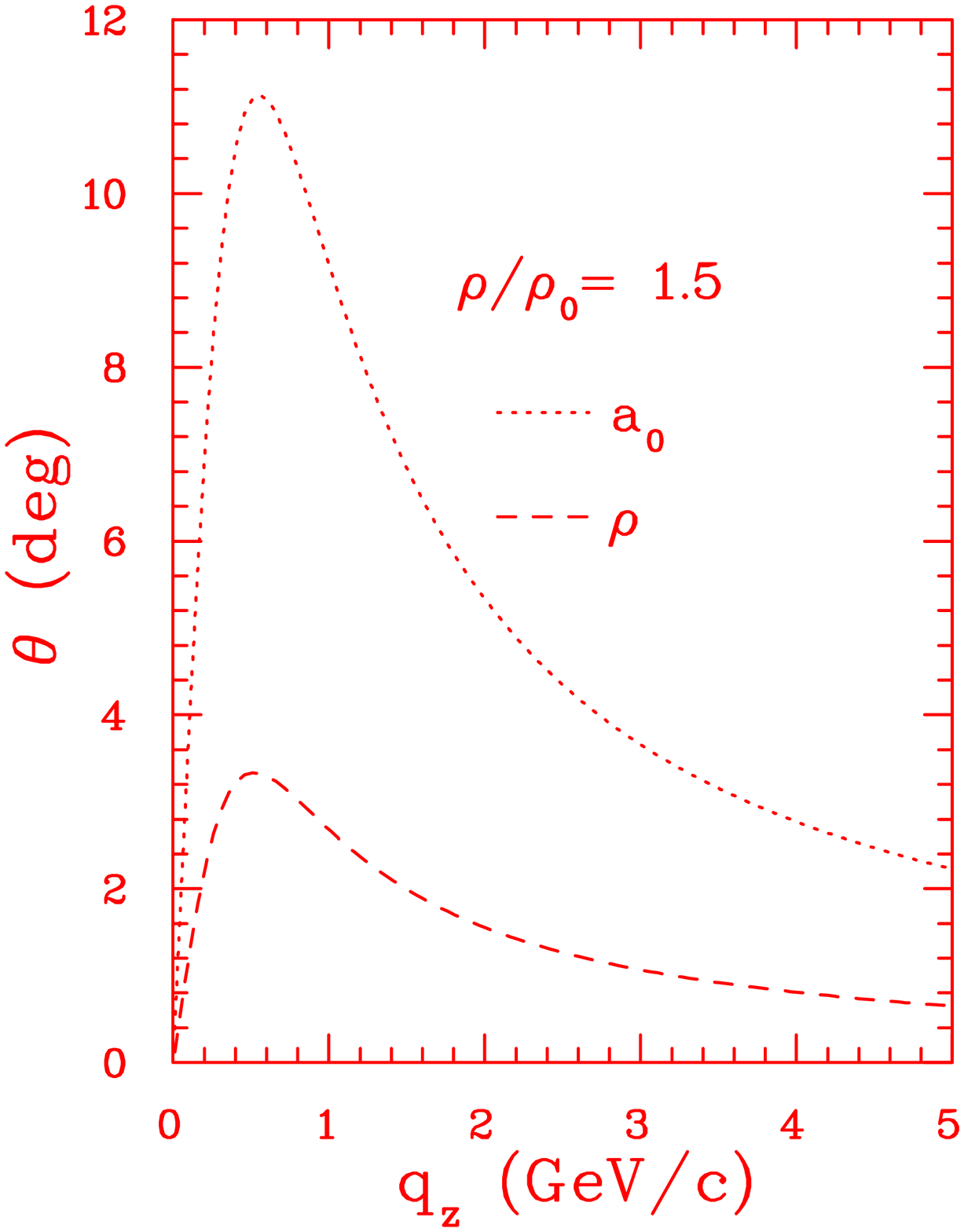,height=6.0cm,angle=0}
\end{center}
 \caption[Dispersion curve] 
{\small The dispersion curve and mixing angle at 
${\var\rho}$=2.5${\var\rho}_0$. \label{fig:dispersion}}
\end{figure}
The left panel of Fig.~\ref{fig:dispersion} shows the 
relevant dispersion curves with
and without mixing at density ${\var\rho}$=2.5${\var\rho}_0$. 
As only the L mode mixes with the scalar mode, we do not consider the
T mode. The later in fact is the same as presented in Ref. \cite{abhee97} for
the $\rho$ meson, and in Ref. \cite{saito98} for $\sigma - 
\omega$ mixing.  
The effect of mixing on the pole masses, as evident from
Fig. \ref{fig:dispersion}, are found to be small. However, the mixing 
could be large when the mesons involved go off-shell. 

To calculate the mixing angle, one diagonalises the mass matrix\cite{abhee97} 
\begin{figure}[h!]
\begin{center}
\epsfig{file=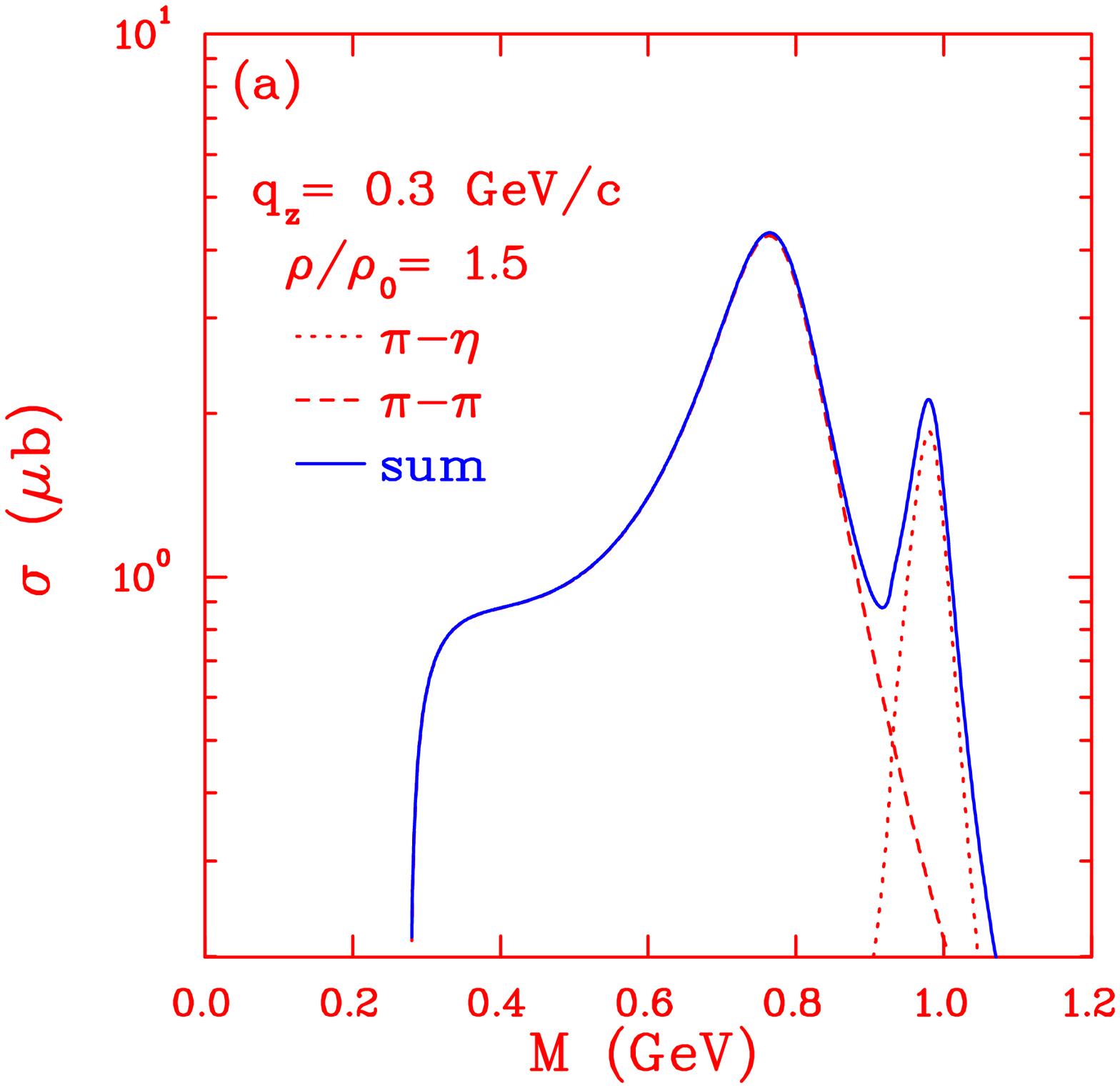,height=6.0cm,angle=0}
\epsfig{file=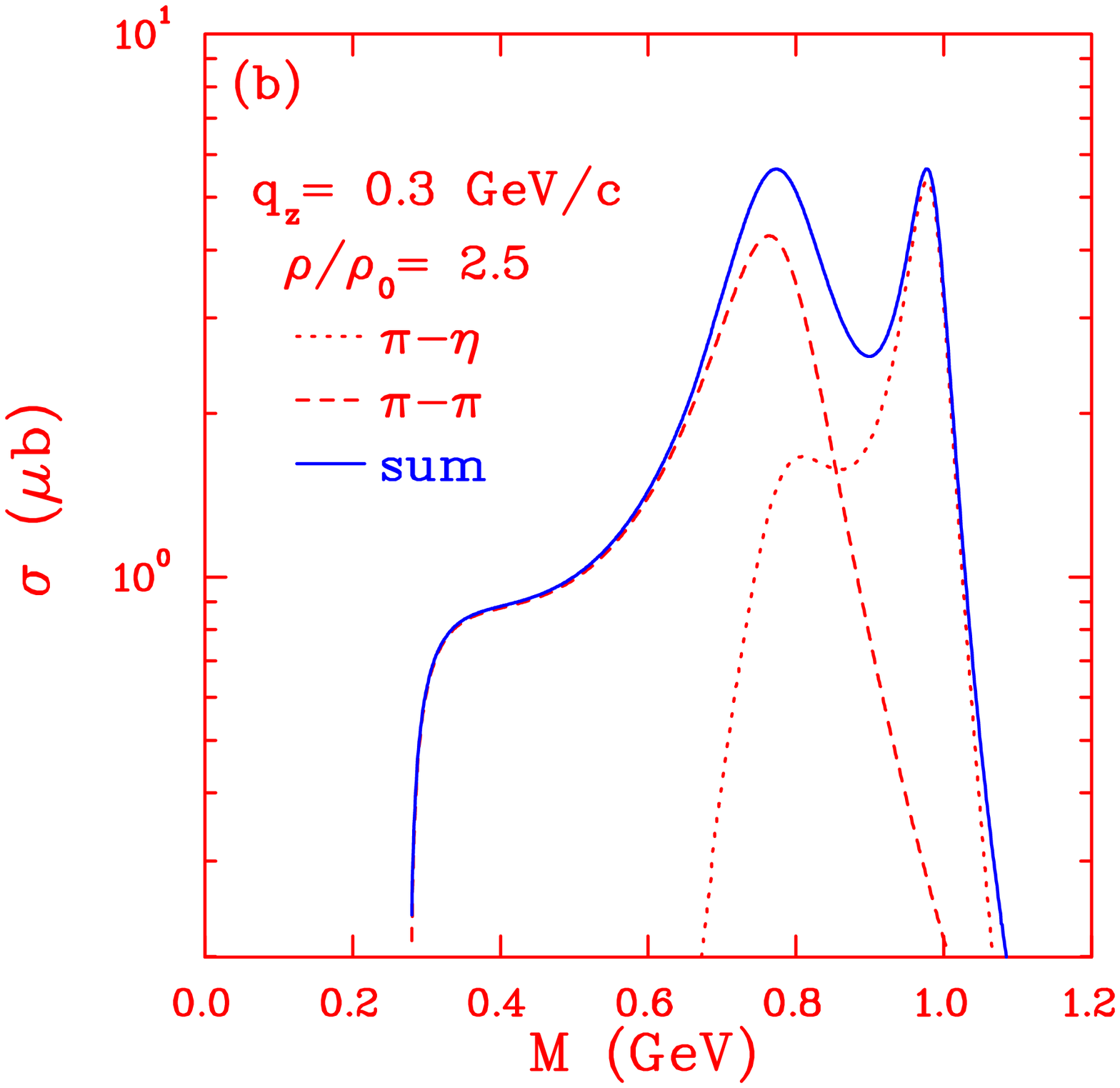,height=6.0cm,angle=0}
\end{center}
 \caption[Dilepton spectrum] 
  {\small Dilepton production cross section for 
$\pi + \eta \rightarrow e^+ + e^-$ through matter-induced $\rho-{a_0}$ 
mixing, and for  $\pi + \pi \rightarrow e^+ + e^-$.
 \label{fig:dilepton}}
\end{figure}
with the mixing and obtains
\bea
\theta_{mix} = \frac{1}{2} \arctan (\frac{2 \Pi^{\rho{a_0}}_{mix}}{
m_{a_0}^2 - m_\rho^2 - \Pi_L^\rho + \Pi^{a_0} })
\label{mixangle}
\eea
In Eq.~(\ref{mixangle}) $\Pi^{\rho {a_0}}_{mix}= M_i/|{\vec q}|\Pi_0$ which
increases with density. $\Pi_0$ is the zero'th component of 
Eq.~(\ref{mixamp}).  
The momentum dependence for a density of
1.5 times higher than the normal nuclear matter density is shown in
the right panel of Fig. \ref{fig:dispersion}. 
This shows that for momenta beyond 
$|{\vec q}|\approx 0.2 $ GeV/c the mixing is quite appreciable. It should
also be noted that the mixing angle vanishes at
$|{\vec q}|$ = 0 as it should.

The $\rho$-$a_0$ mixing opens a
new channel, $\pi + \eta \rightarrow e^+ + e^-$, in dense
nuclear matter through n-n excitations .
The cross-section for this
process is expressed in terms of the mixing amplitude ($\Pi_0$)
\cite{tdmg} and is plotted in Fig.~\ref{fig:dilepton}. 

One can notice in Fig. \ref{fig:dilepton} that the  process, 
$\pi + \eta \rightarrow e^+ + e^-$, at densities higher than ${\bf \rho_0}$,
not only enhances the overall production of lepton pairs but also
induces an additional peak near the $\phi$ mass region. The contribution
at the $a_0$ mass is comparable to that of  
$\pi + \pi \rightarrow e^+ + e^-$ near the $\rho$ peak, 
for densities higher than ${\bf \rho_0}$. 
Fig.~\ref{fig:dilepton} also shows that as the density goes even higher the
dilepton yield arising out of the mixing also increases further. 

\begin{figure} [htb!]
\begin{center}
\epsfig{file=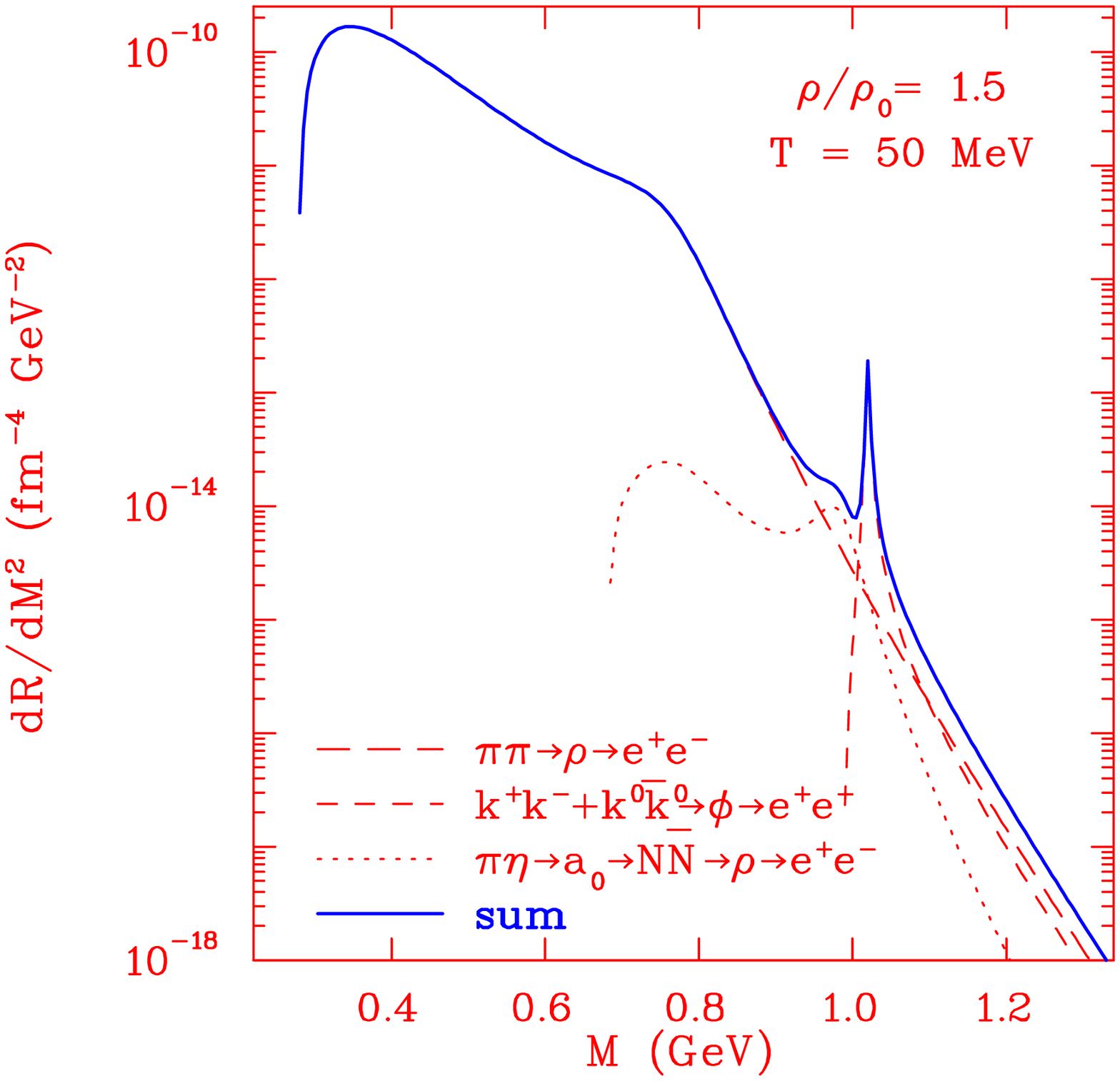,height=6cm,angle=0}
\epsfig{file=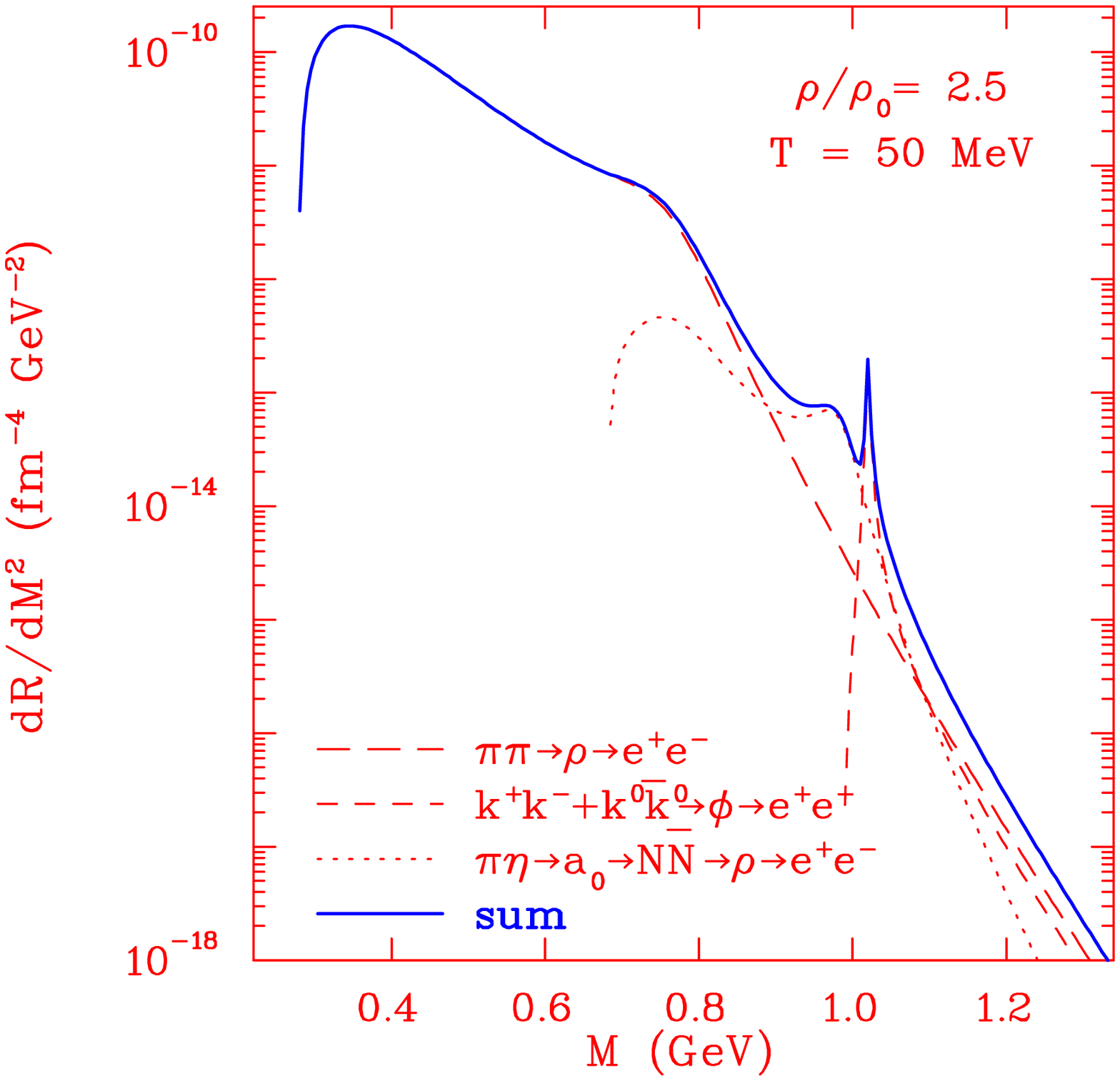,height=6cm,angle=0}
\end{center}
 \caption[rates]
  {\small Dilepton production rates for T=50 MeV.\label{fig:rates}}
\end{figure}
Further studies are in progress to assess finite temperature effects and to
incorporate the necessary many-body machinery. We will show here only some 
preliminary results for the dilepton production rate including the 
effect of $\rho-{a_0}$ mixing. We work the independent particle approximation 
of kinetic theory. We assume a thermal gas of mesons at $T=50 MeV$. 
Fig. ~\ref{fig:rates} shows the dilepton production rate as compared with 
the standard $\pi \pi$ and $K {\bar K}$ annihilation.  
One can observe from Fig \ref{fig:rates}
that even at density $\rho/\rho_0=1.5$ the contribution of the mixing 
wins over the $\pi$-$\pi$ annihilation rate in the vicinity of M = 1.0 GeV. 
Naturally at higher density this goes up as evident from the right panel
of Fig. \ref{fig:rates}. 

We have highlighted the possibility of 
$\rho$-$a_0$ mixing in dense nuclear matter. 
We observe the appearance of an additional peak at a dilepton invariant
mass that corresponds to that of the $a_0$. With sufficient
experimental resolution, this effect could be observed. Maybe not as
an individual peak, but probably more realistically as a shoulder in 
the $\phi$
spectrum. This feature would then be exclusively temperature 
and density-dependent and would thus be the reflection of a genuine
in-medium effect.


This work was supported in part by the Natural Sciences and Engineering
Research Council of Canada and in part by the Fonds FCAR of the 
Qu\'ebec Government.

\end{document}